**Embracing Agile methodology during DevOps Developer Internship Program**

Amol S Patwardhan, Jon Kidd, Tiffany Urena, Aishwarya Rajgopalan.

**Introduction:** The DevOps team adopted agile methodologies [1] during the summer internship program as an initiative to move away from waterfall. The DevOps team implemented the Scrum software development strategy [2] to create an internal data dictionary web application. This article reports on the transition process [3] and lessons learned from the pilot program.

**Waterfall methodology:** For several decades the development teams have used waterfall process [4] for implementing software. But in the recent years several software companies, technical articles and research papers have acknowledged the benefits of agile methodologies over waterfall and have attempted to adopt agile processes. Software industry is constantly trying to keep up with competition to deliver products to market in a short duration, while also maintaining quality and meeting customer expectations and adopting new technologies at the same time. Vendors facilitating the switch to agile provide plenty of presentations about advantages of agile but lack definitive steps detailing how an organization can change the deeply rooted processes and overcome resistance to change. The transition needs to occur from within an organization and the DevOps [5] team decided to take the first step towards going agile.

**DevOps:** The DevOps team serves as facilitator between the software development team and the information technology (IT) operations team. The DevOps team at our company had an opening to hire a developer intern for the summer. The team saw an opportunity to use the internship as a stepping-stone for adopting agile in development process. The DevOps team worked on projects with smaller scopes, internal IT projects, utility tools, automation, infrastructure projects and non-production as well as production issues support. All of these projects could server as potential candidate for embracing agile methodology. The following benefits were identified in moving to agile methodology during the internship program. 1) The process would provide a real world example for assessing the advantages of going agile within the organization. 2) The process would provide a fulfilling internship experience. It would also inculcate agile mentality early in the career for the intern. 3) The process would help identify the hurdles in transitioning from waterfall to agile.

**Project Overview:** The core product at our company is a web application built on micro services architecture. There are several relational databases used behind the scene with very little documentation about the database, tables, columns and stored procedures. There was a need to create a centralized, internal web site containing information about all the databases. On completion the website was hosted in one of the non-production environments. New developers, Business Analysts and QAs had been finding it difficult to understand the intent, naming convention and business use case behind a certain database, table and columns and their cryptic names did not provide much information. Some common questions asked were: What did a specific database, table, column name mean? What was the rationale behind the name? What is it used for? What are the standards for data types of columns? DevOps chose the data dictionary project as the candidate for the agile initiative. Since there was no formal documentation available to describe these database objects internally in IT departments, the data dictionary project was started to serve as a self-contained project with sufficiently simple/small scope for a DevOps developer intern. Scrum strategy was chosen to practice the agile methodology as part of the pilot initiative.

**Scrum Roles:** The various roles for the Scrum methodology were defined as follows:

a) Scrum team: The scrum team consisted of DevOps manager, business analyst, and senior software engineer and developer intern.
b) Scrum master: The DevOps manager served as the Scrum master. His responsibilities were to remove any impediments to the progress of the project. For instance, the intern's development machine started crashing in the second week at the job. The scrum master immediately raised a support ticket with the desktop support team. He followed up with the technician to get the issue resolved and ensured that the coding effort was not affected. A replacement laptop was arranged in the interim to ensure that there was no downtime. He also acted as the facilitator of team events to ensure regular progress. For example, the scrum master organized a meeting and initiated an email-communication between subject matter experts and the developer intern so that information about the databases could be shared and added to the database dictionary.
c) Product Owner: The business analyst on the DevOps team acted as the Product Owner. Since the project was an internal IT initiative, the product owner was also the mock customer. The Product owner provided important feedback about the language, text, color scheme, layout and expected functionality from the Data Dictionary project. The product owner played an important role in arranging milestone meetings, limiting the scope of the project and helped manage communication between various stakeholders across departments keen on using the data dictionary website and the DevOps development team.
d) Development team: The development team for the data dictionary project consisted of senior computer analyst, senior software engineer and the development intern. Both the resources belonged to the DevOps team with senior analyst mostly assisting with the architecture and design of the data dictionary web application. The technology chosen for implementation was ASP.NET MVC using C# and MS SQL as the backend system. The senior software engineer served as a technical mentor for the developer intern.

**Planning:** The scrum team met and outlined the sprint duration (4 weeks = 20 business days) for implementing the data dictionary project. The team defined the definition of done/complete project as the day the project would receive sign off for all the stories from quality assurance team.

**Daily stand ups:** A daily stand up at 9 am was organized. The scrum master and the development team met and shared status updates. The three things that were discussed during the daily stand ups were as follows: 1) task completed by the developer intern, 2) task goals for the day to be completed by the developer intern and 3) any challenges experienced while implementation of each tasks. The summary of daily stand up was emailed to the whole group and served as the daily stand up status board.

**Coding, unit testing, and check-ins with TFS:** The coding was done using out of the box project template available in Visual Studio 2015 for web application using ASP.NET MVC in C#. The webpages, UI layout, user stories, classes, objects and business entities were designed and discussed on a white board. Critical logic in the code was documented inline and functions were documented at the top.

**Pair programming benefits:** To further embrace agile methodology one of the extreme programming practice of pair programming [6] was adopted. The pair programming sessions were limited to half an hour each day. In each session the senior programmer would act as the driver and the intern served as the navigator. Once the knowledge about coding best practices and internal coding standards was transferred the developer intern was provided opportunity to implement the functionality inside the code stubs/snippets created by the senior programmer.

**Code review:** Even though pair programming served as a real time code review, there was still a need to look at the code completed by the intern at the end of the day. This was achieved by daily code reviews performed by the senior software engineer. The feedback allowed the intern to make corrections in the code in a timely manner.

**Documentation:** The comments added to the code using the summary feature of the ASP.NET allowed automated generation of documentation with every build.

**Testing and Feedback:** Comprehensive unit tests were written to check various code paths in the methods inside the controllers. Check-ins to TFS and code reviews were performed daily to ensure the code quality was maintained. The product owner was sent emails with screenshots asking for feedback. This allowed timely response and clarification from any misunderstandings regarding any user story.

**Continuous deployment:** The deployment was handled using custom built continuous automated deployment tool built using TFS API, web API, angularJS and powershell scripts. The deployment was enabled by the DevOps build engineer. Using the TFS builds and automated deployment tool, the intern was able to deploy the project daily and test the web application code in an actual quality assurance (QA) environment instead of developer machine.

**Burn down chart:** The figure below shows how the daily stand ups allowed to make regular progress which could be closely monitored and compared with ideal effort using the burn down chart.

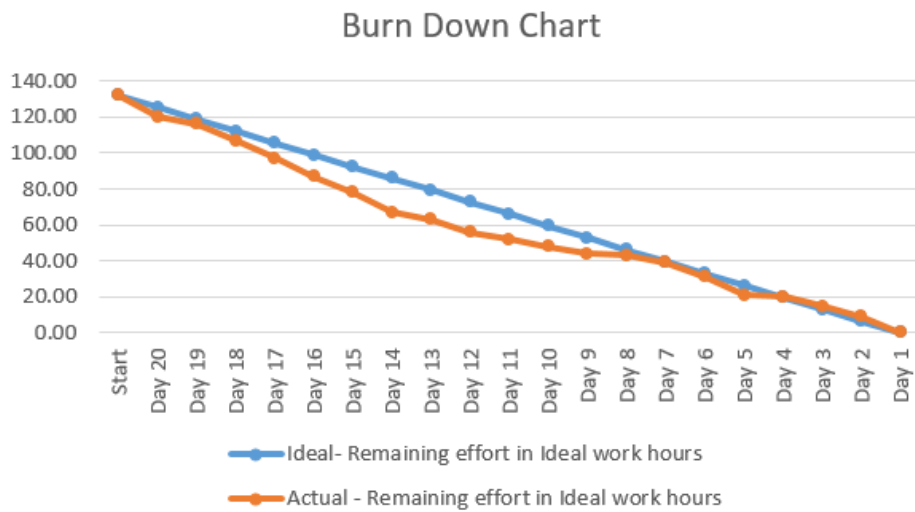

**Intern questionnaire:** The intern was asked a series of feedback questions and the response to all the questions were positive. 1) Was the internship helpful in learning day-to-day functioning of an IT department? 2) Was the internship useful in learning new web development technologies? 3) Was the pair programming experience helpful? 4) Did agile methodology provide opportunity to receive timely feedback and delivering the tasks on time? 5) Was the internship fulfilling? 6) Were the mentor and manager helpful? 7) Do you have a better understanding of agile methodologies 8) Would this experience help you in future job profiles?

**Conclusion:** The pilot project to adopt agile methodologies using a DevOps developer internship program was a great success in terms of:

1) Proving benefits of Scrum to the management.
2) Inculcating agile mentality early in the professional career of a software developer intern.
3) Implementing an internal web application and learning SCRUM while working real time on the project.
4) The daily standups and email updates helped track the progress.
5) Immediate resolution of technical issues helped make regular progress with minimum down time.
6) Once the project was complete the participants felt comfortable adopting agile on a larger scale.
7) The transition was smoother because all the team members were equally enthusiastic about agile methodologies and even though the product owner was a remote employee, the status updates, meetings using screen share, email and conference calls proved to be adequate tools to overcome the hurdle of long distance communication.